\documentclass[submission, Proceedings]{SciPost}

\hypersetup{
    colorlinks,
    linkcolor={red!50!black},
    citecolor={blue!50!black},
    urlcolor={blue!80!black}
}

\usepackage[bitstream-charter]{mathdesign}
\usepackage{particlessymb}
\usepackage{caption}
\usepackage{subcaption}
\DeclareSymbolFont{usualmathcal}{OMS}{cmsy}{m}{n}
\DeclareSymbolFontAlphabet{\mathcal}{usualmathcal}

\newcommand{\delphes}{\textsc{Delphes}\xspace}
\newcommand{\whizard}{\textsc{Whizard}\xspace}

\newcommand{\figwidth}{0.50\textwidth}
\newcommand{\twofigwidth}{0.45\textwidth}

\begin{document}

% TODO: write your article's title here.
% The article title is centered, Large boldface, and should fit in two lines
\begin{center}{\Large \textbf{
Searches for invisible scalar decays at CLIC
}}\end{center}

% TODO: write the author list here. Use initials + surname format.
% Separate subsequent authors by a comma, omit comma at the end of the list.
% Mark the corresponding author with a superscript *.
% \begin{center}
%   Krzysztof Mekala\textsuperscript{1$\star$}, 
%   Aleksander Filip \.Zarnecki\textsuperscript{1},
%   Bohdan Grzadkowski\textsuperscript{1} and
%   Michal Iglicki\textsuperscript{1}
% \end{center}

\begin{center}
  Krzysztof Mekala\textsuperscript{$\star$}, 
  Aleksander Filip \.Zarnecki,
  Bohdan Grzadkowski and
  Michał Iglicki
\end{center}

% TODO: write all affiliations here.
% Format: institute, city, country
\begin{center}
%{\bf 1} 
Faculty of Physics, University of Warsaw, Warsaw, Poland
\\
% TODO: provide email address of corresponding author
* k.mekala@student.uw.edu.pl
\end{center}

\begin{center}
\today
\end{center}

% For convenience during refereeing (optional),
% you can turn on line numbers by uncommenting the next line:
% \linenumbers
% You should run LaTeX twice in order for the line numbers to appear.

\definecolor{palegray}{gray}{0.95}
\begin{center}
\colorbox{palegray}{
  \begin{tabular}{rr}
  \begin{minipage}{0.1\textwidth}
    \includegraphics[width=22mm]{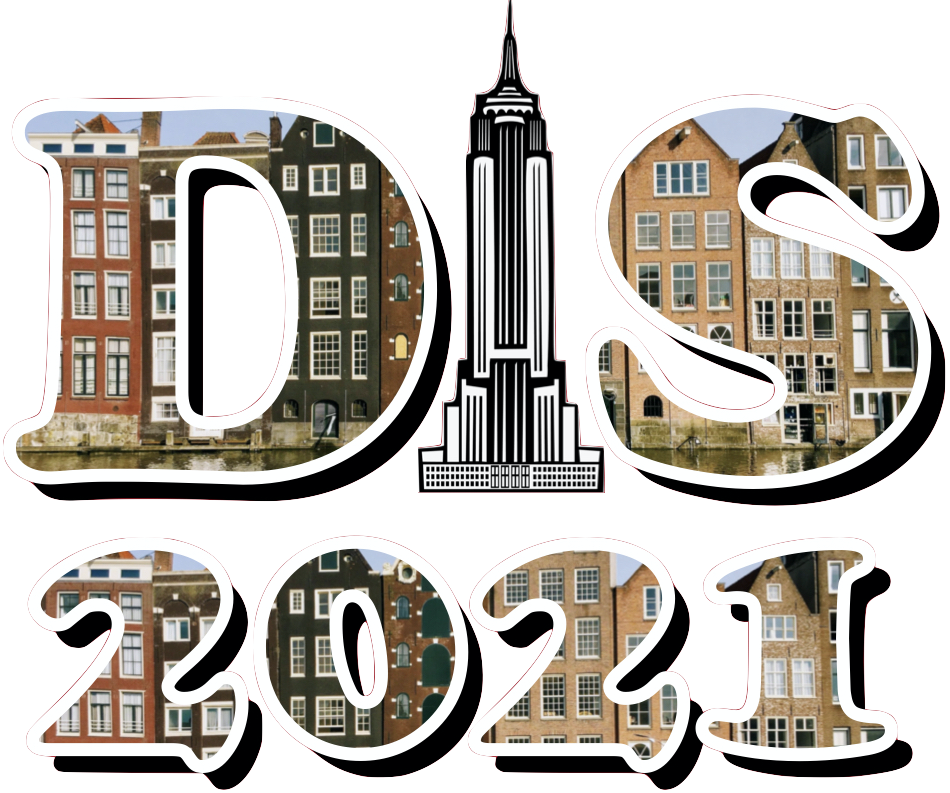}
  \end{minipage}
  &
  \begin{minipage}{0.75\textwidth}
    \begin{center}
    {\it Proceedings for the XXVIII International Workshop\\ on Deep-Inelastic Scattering and
Related Subjects,}\\
    {\it Stony Brook University, New York, USA, 12-16 April 2021} \\
    \doi{10.21468/SciPostPhysProc.?}\\
    \end{center}
  \end{minipage}
\end{tabular}
}
\end{center}

\section*{Abstract}
{\bf
The Compact Linear Collider (CLIC) is a proposed TeV-scale
high-luminosity electron-positron collider at CERN.
The first CLIC running stage, at 380\,GeV, will focus on precision Higgs boson
and top quark studies while the main aim of the subsequent high-energy stages, at
1.5\,TeV and 3\,TeV, is to extend the sensitivity of CLIC to different Beyond
the Standard Model (BSM) scenarios.

We studied the prospects for measuring invisible Higgs boson and
additional heavy scalar decays using CLIC data at
380\,GeV and 1.5\,TeV. The analysis is based on the \whizard event
generator, with fast simulation of the CLIC detector response parametrised by the \delphes package.
We present the expected limits for the invisible decays of the 125 GeV
Higgs boson, the cross section limits for production of an
additional neutral Higgs scalar, assuming its invisible decays, and
limits on the mixing angle between the SM-like Higgs boson and the new
scalar of the "dark sector" in the framework of the vector-fermion
dark matter model. 
}

% TODO: include a table of contents (optional)
% Guideline: if your paper is longer that 6 pages, include a TOC
% To remove the TOC, simply cut the following block
%\vspace{10pt}
%\noindent\rule{\textwidth}{1pt}
%\tableofcontents\thispagestyle{fancy}
%\noindent\rule{\textwidth}{1pt}
%\vspace{10pt}

\section{Introduction}
\label{sec:intro}

The Higgs boson of the Standard Model (SM) with a mass of about
125\,GeV is expected to decay into a wide variety of final states, 
including unobservable ones such as two $Z$ bosons decaying to neutrinos.
A larger branching fraction to unobservable final states is predicted 
in many extensions of the Standard Model. % Consider citation here !!!  
Experimental constraints on the invisible Higgs boson decays can be set either directly,
by searching for such decays in channels where Higgs boson production
can be tagged independently of the decay mode e.g.\  via vector boson
fusion in $pp$ collisions or associated production with a $Z$ boson in
$e^{+}e^{-}$ collisions, or indirectly, based on a global fit to
all production and decay channel measurements, assuming that the total
width of the Higgs boson can also be directly measured. 
As of today, the best direct limits on invisible Higgs boson decays come
from experiments at the LHC -- at 95\% C.L. upper limits on the branching fraction is
less than 13\% for ATLAS~\cite{ATLAS:2020cjb} and 19\% for
CMS~\cite{CMS}.

In this analysis, we studied the prospects of measuring
invisible Higgs decays at CLIC with 380\,GeV and 1.5\,TeV data in the
Higgs-strahlung process ($e^{+}e^{-} \to ZH$)~\cite{Mekala:2021ftq}. 
The sensitivity is dominated by the hadronic $Z$ boson decay channel
providing an order of magnitude higher observed yield
\cite{Thomson:2015jda,Bambade:2019fyw}. 
Furthermore, we extend our search for invisible decays of the SM-like
Higgs boson to the search for production and invisible decays of
another scalar particle, $H'$, with an arbitrary mass. 
We then interpret our results in terms of the limits on the scalar
sector mixing angle in the Higgs-portal scenarios.
For illustration, we employed the vector-fermion dark matter model
(VFDM) \cite{vfdm1,vfdm2}, an extension of the Standard Model 
with one extra scalar, two Majorana fermions and one gauge boson.

\section{Event generation and detector simulation}

The results are based on a fast simulation of the CLICdet \cite{clicdet}
detector response provided by the parametric \delphes framework \cite{delph}.
Control cards prepared for the new detector
model CLICdet \cite{delcards} were modified to make Higgs particles
`invisible' in the simulation (ignored when generating detector
response), so that the invisible scalar decays can be modeled by
defining the Higgs boson as stable. 
Signal and background event samples were generated using \textsc{Whizard} 2.7.0
\cite{whiz1, whiz2}, using the beam energy profile
expected for CLIC running at 380\,GeV and 1.5\,TeV.
The $e^{+}e^{-} \to ZH'$ process, where the Higgs-like scalar is 
produced (with decay into an invisible final state) together with a
$Z$ boson decaying into a quark-antiquark pair was considered as the signal.
Masses of the new scalar in the range 120--280\,GeV (for the first stage of CLIC) and
150--1200\,GeV (for the second stage) were considered.

As the background, we studied processes both with and without
Higgs boson production.
We also took into account possible background contributions from hard
$\gamma\gamma$ and $e^{\pm}\gamma$ interactions, where we included beamstrahlung 
photons, as well as photon radiation by the incoming electrons, as described
by the Equivalent Photon Approximation.

For 380\,GeV collisions, two running scenarios are
considered: a baseline scenario with an integrated luminosity of
1000\,fb$^{-1}$ \cite{gHZZ} and an extended one with 4000\,fb$^{-1}$
of data collected at the first stage \cite{Robson:2020lhl}. 
As the same integrated luminosity is expected for both
electron beam polarisations, the data can be considered as
unpolarised in the combined analysis.
At 1.5\,TeV,  the two electron beam polarisations are considered separately,
with 2000\,fb$^{-1}$ collected with --80\% polarisation and
500\,fb$^{-1}$ with +80\% polarisation \cite{gHZZ}.

\section{Data analysis}

Only events with the expected signal
signature, two reconstructed jets with an invariant mass corresponding
to the mass of the $Z$ boson and no other activity in the detector,
were accepted at the preselection stage.
In particular, all events with reconstructed isolated leptons
(electrons or muons) or isolated energetic photons were excluded from
the analysis.  
Quantities describing event topology were then considered. 
First, the distributions of resolution parameters associated with the VLC
jet clustering algorithm were analyzed. 
The parameters y$_{23}$ and y$_{34}$ were used to suppress events with higher jet
multiplicities.
Only events for which y$_{23}<0.01$ and y$_{34}<0.001$  were considered.
After forcing the event into a two-jet topology using the same algorithm, 
the invariant mass of the two-jet final state, m$_{jj}$,
was also required to be consistent with the mass of the $Z$ boson so
only events with 80\,GeV$<$m$_{jj}<$100\,GeV were selected for
further analysis.
A fiducial requirement,
$|\cos(\theta)| > 0.8$, where $\theta$ is the polar angle of the reconstructed dijet, 
was made to exclude events produced close to the beam axis, where background dominates. 

\begin{figure}[tb]
  \centerline{\includegraphics[width=\figwidth]{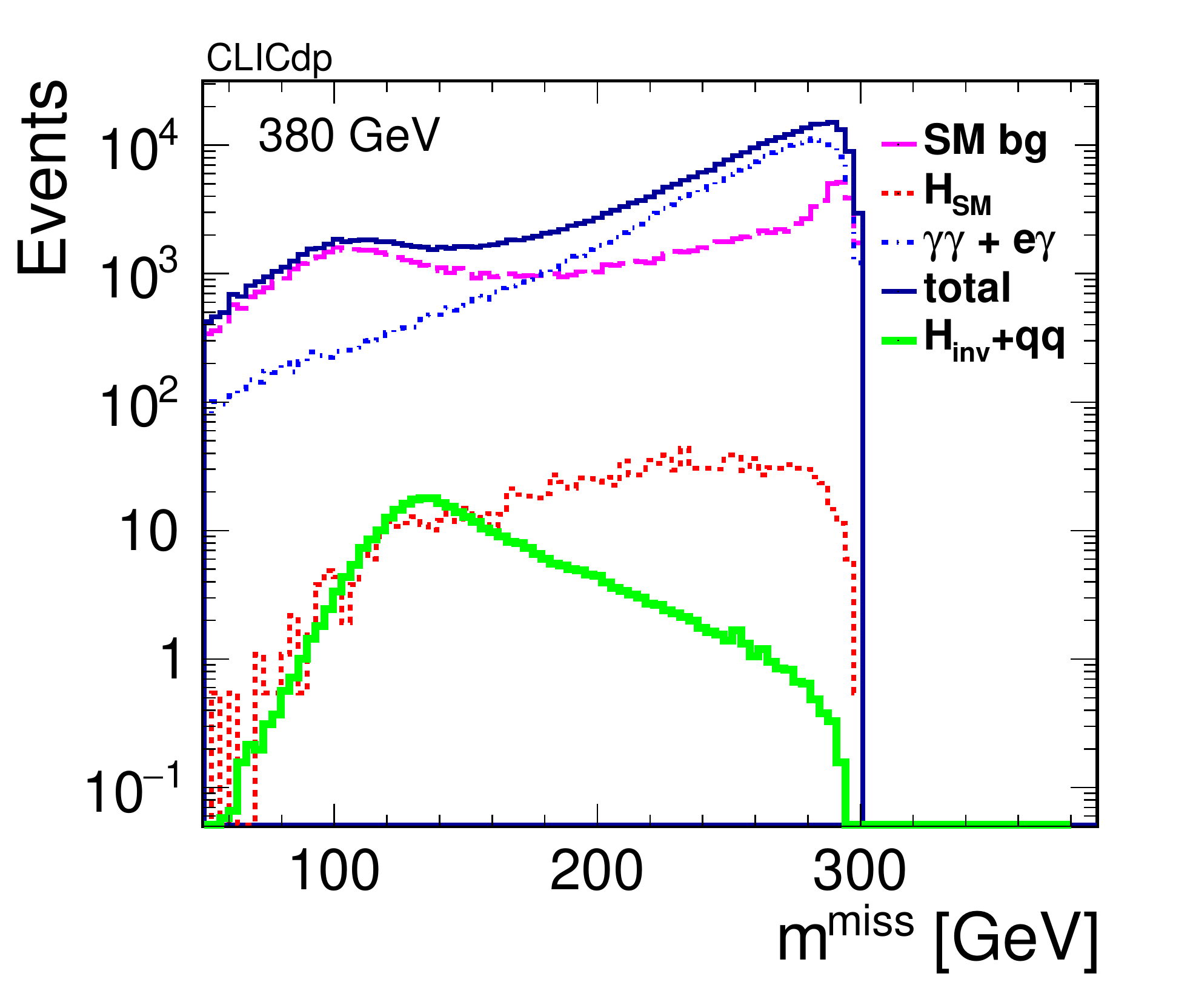}}
  \caption{Reconstructed invariant mass of the products of the invisible Higgs boson decay expected for different event
    samples after preselection cuts:
    total background (thin solid black line), $e^{+}e^{-}$ background
    without Higgs boson production (long-dashed pink line), background
    of SM Higgs boson production and decays (short-dashed red line),
    photon interactions (dash-dotted blue line) and signal (thick solid
    green line), assuming
    integrated $e^{+}e^{-}$ luminosity of 1000 fb$^{-1}$ collected at 380\,GeV. 
    The signal sample is normalised to BR$(H \to inv)=1\%$.
}
  \label{fig:mrec_sm}
\end{figure}

Figure~\ref{fig:mrec_sm} shows the expected distribution of the
invariant mass of the invisible final state inferred from 
energy-momentum conservation for CLIC at 380\,GeV after the preselection cuts. 
For the background sample, the distribution has two maxima: one at around
300\,GeV, which is the kinematic limit (as we require two
jets to have an invariant mass of at least 80\,GeV) and the second one at around
90\,GeV, which is mainly due to on-shell invisible $Z$ boson decays. 
For signal events, with the cross section normalised in the Fig.~\ref{fig:mrec_sm} to BR$(H \to
inv)=1\%$, the expected recoil mass distribution is consistent with
the SM Higgs boson mass of 125\,GeV.

In the final analysis stage a Boosted Decision Trees (BDT) algorithm,
as implemented in TMVA framework\cite{TMVA}, was used for event
classification, with five input variables: dijet energy, dijet
invariant mass, reconstructed recoil mass, missing transverse momentum
and angle between the two reconstructed jets in the laboratory frame. 
The cut on the BDT response was then selected to give the highest
expected significance for the signal observation.
For invisible decays of the 125\,GeV Higgs boson at 380\,GeV, a
BDT response cut of about 0.14 was used, corresponding to a signal selection efficiency of about 50\% and background
rejection efficiency of about 95\%. 
The same analysis procedure was applied to signal and background samples
generated at 1.5\,TeV, separately for two considered
electron beam polarisation settings. The purpose was to estimate the expected sensitivity
of CLIC experiment to production and invisible decays of a new scalar state.

\section{Results}

For the 380\,GeV operation, assuming that the measured event
distributions are consistent with the predictions of the Standard
Model and that systematic uncertainties are small relative to statistical uncertainties, the expected 95\% C.L. limit
 on the invisible branching ratio of the 125\,GeV Higgs coming from the BDT analysis is:  
\begin{center}
BR$(H \to inv)<1.0\%  \;\;\; (0.5\%)$
\end{center}
for an integrated luminosity of 1000\,fb$^{-1}$ (4000\,fb$^{-1}$).
The discovery of a new decay channel at $5 \sigma$ level (and therefore also of new, invisible
particles) is possible for an invisible Higgs boson branching ratio above 3.0\% (1.5\%).

\begin{figure}[tb]
\centering
\begin{subfigure}[b]{\twofigwidth}
  \includegraphics[width=\textwidth]{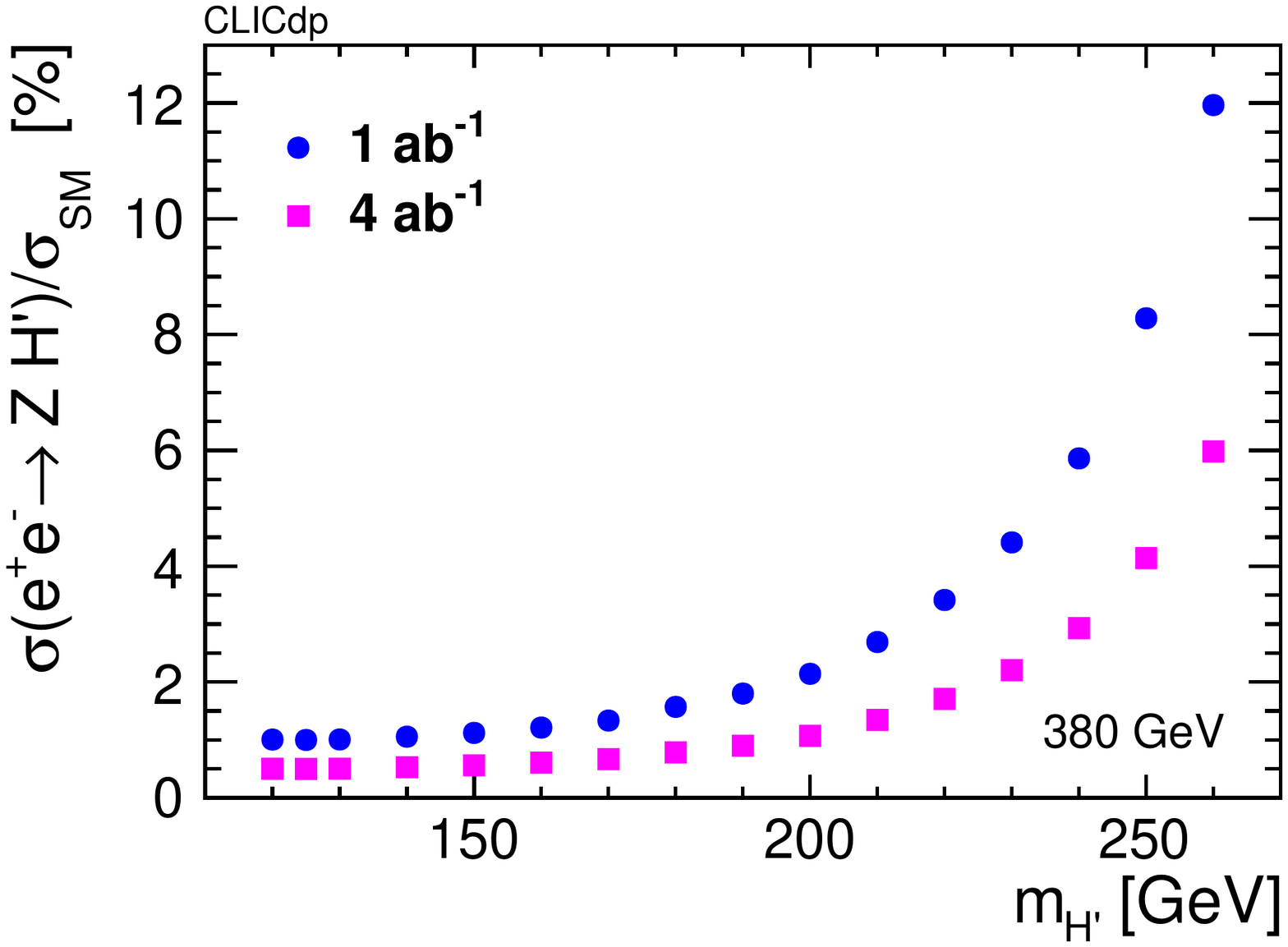}
  \caption{380 GeV}
  \end{subfigure}
\begin{subfigure}[b]{\twofigwidth}
  \includegraphics[width=\textwidth]{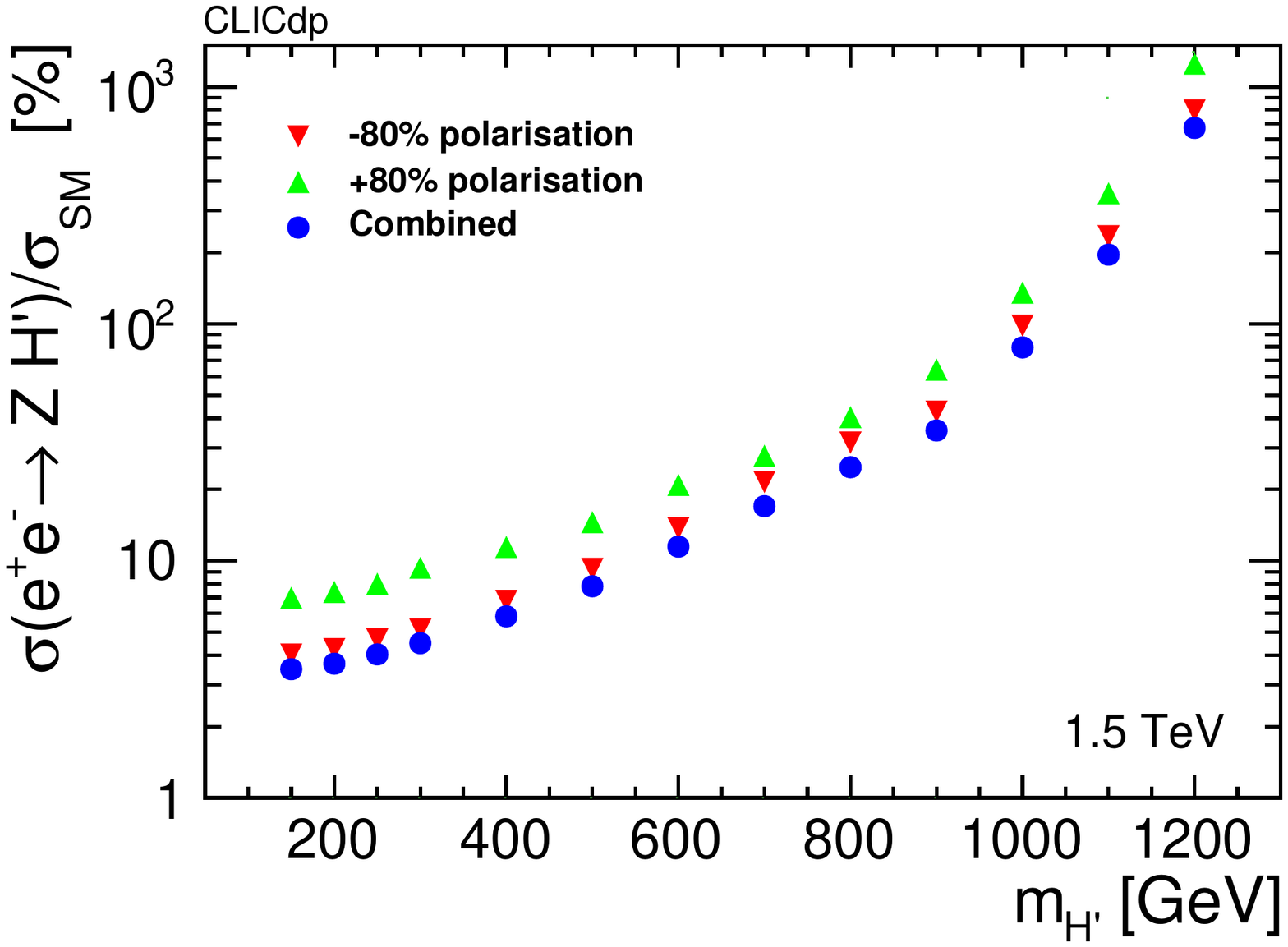}
  \caption {1.5 TeV}
  \end{subfigure}
    \caption{Expected 95\% C.L. limits on the production cross section of the
      new scalar $H'$, relative to the expected SM Higgs production
      cross section, as a function of its mass, for CLIC running at
      380\,GeV (left) and 1.5\,TeV (right). The new scalar is assumed to
      have only invisible decay channels, BR$(H'\to inv) = 100\%$. } 
    \label{fig:limit}
\end{figure}

Figure~\ref{fig:limit} presents the 95\% C.L. limits on the
cross section for the production of the new scalar $H'$ in association
with a $Z$ boson, relative to the expected cross section for the
production of the SM Higgs boson (for a given mass), as a function of
the assumed scalar mass, for 380\,GeV and 1.5\,TeV.
\begin{figure}[tb]
\begin{center}
  \includegraphics[width=\figwidth]{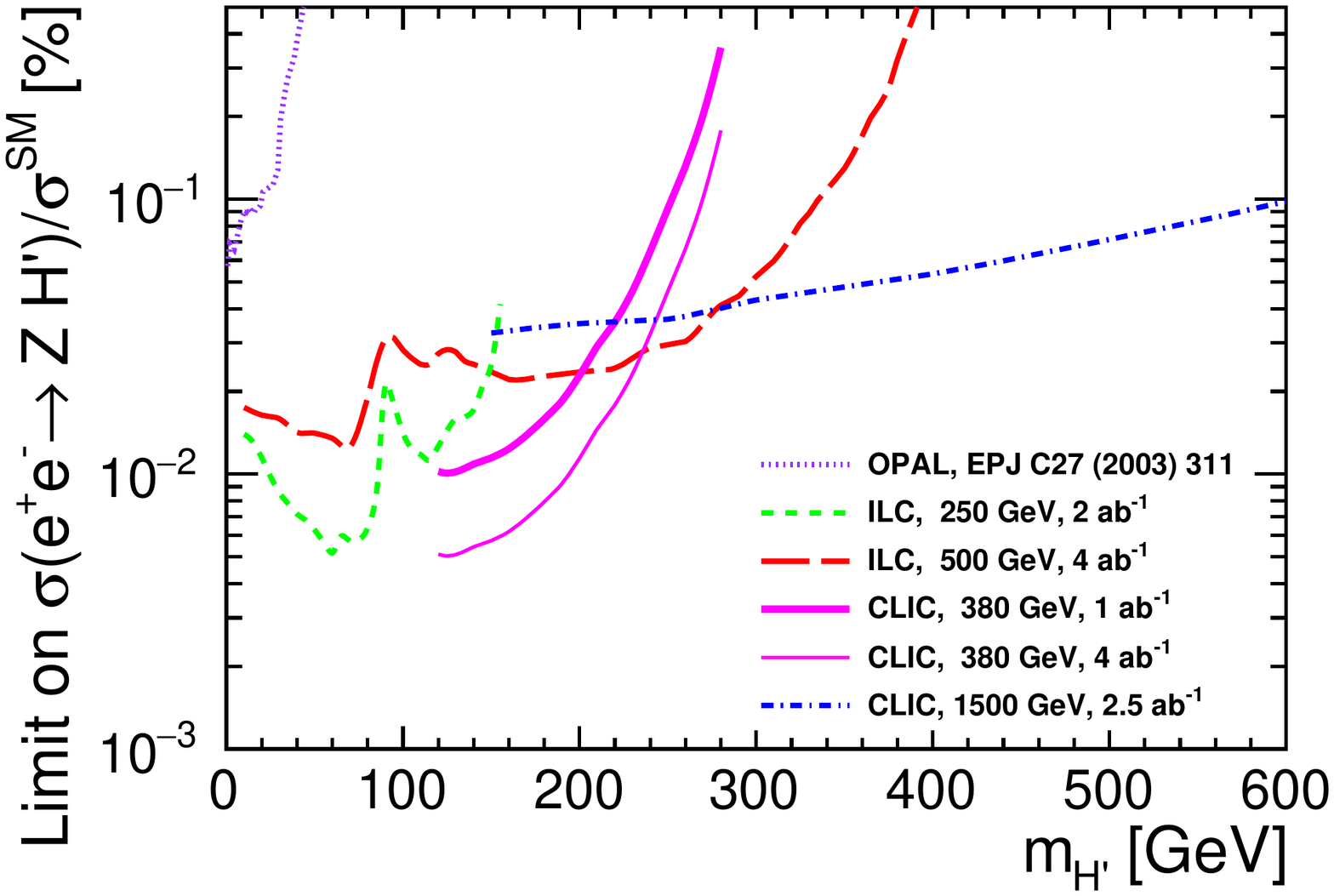}
    \caption{Expected sensitivity of CLIC running at 380\,GeV (thick pink line for 1 ab$^{-1}$ and thin pink line for 4 ab$^{-1}$) and 1.5\,TeV (dotted-dashed blue line) compared
    to the existing limit from LEP (dotted violet line) \cite{Abbiendi:2002qp} and the expected sensitivity of ILC running at 250\,GeV (dashed green line) and 500\,GeV (long-dashed red line) \cite{Wang:2020lkq}. 
    Limits on the production cross section of the
      new scalar $H'$, relative to the expected SM Higgs production
      cross section, are shown as a function of its mass.  For CLIC limits, a new scalar is assumed to have invisible decay channel only, BR$(H'\to inv) = 100\%$, while
      LEP and ILC results are decay-mode independent.} 
    \label{fig:limcomp}
  \end{center}
\end{figure}
In Fig.~\ref{fig:limcomp}, the expected CLIC sensitivity is compared
to the existing limit from LEP \cite{Abbiendi:2002qp} and the expected sensitivity of ILC for 
2000\,fb$^{-1}$ (4000\,fb$^{-1}$) collected at 250\,GeV (500\,GeV) \cite{Wang:2020lkq}. 
The LEP and ILC limits were evaluated in a decay-mode independent
approach, based on the reconstruction of leptonic $Z$ boson decays
($Z\to e^+e^-$ and $Z\to \mu^+ \mu^-$). 

\section{Interpretation}

\begin{figure}[t]
\centering
  \includegraphics[width=\twofigwidth]{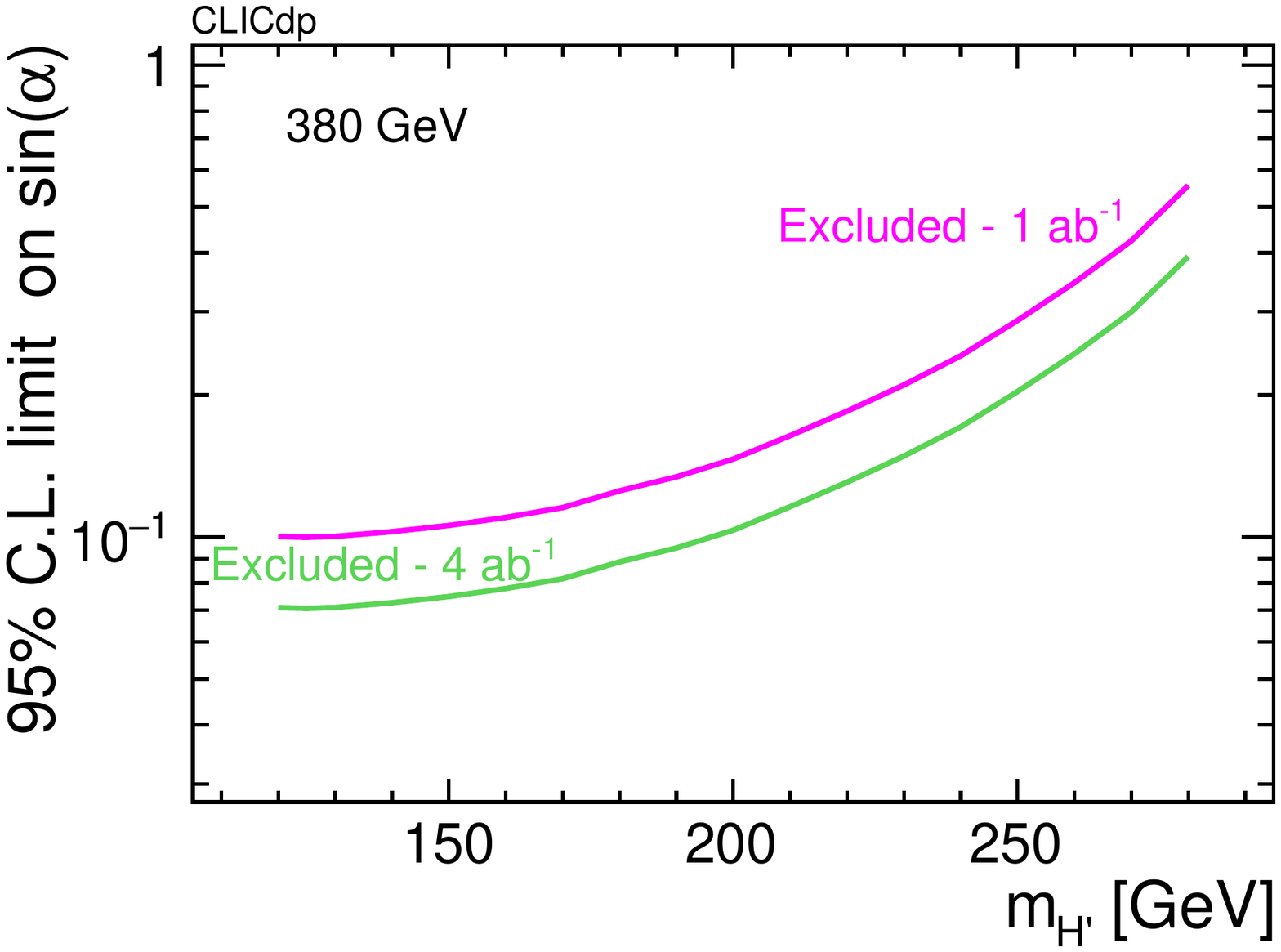}
  \includegraphics[width=\twofigwidth]{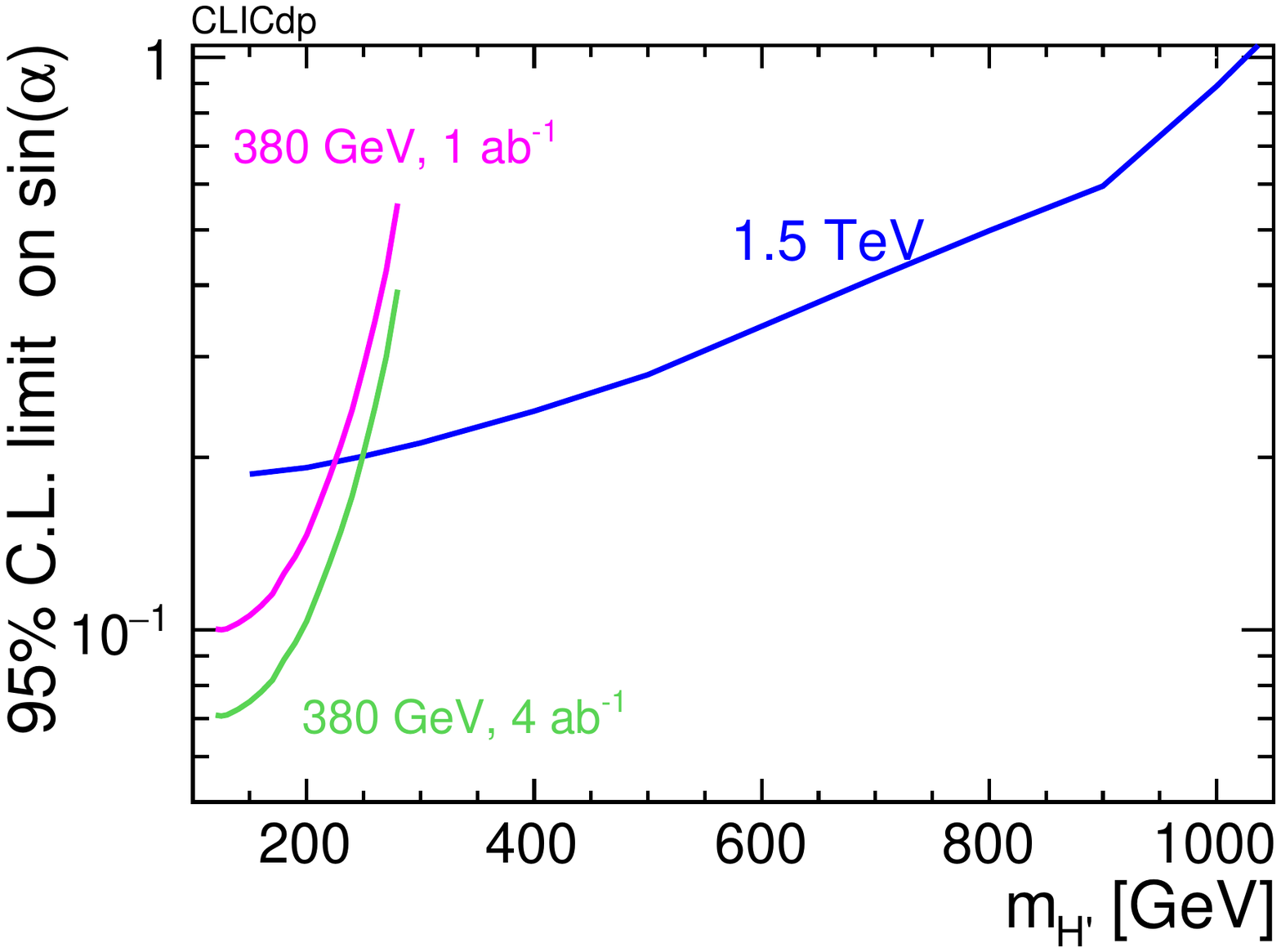}
  
      \caption{Expected limits on the scalar mixing
        angle expressed as a function of the $H'$ mass, for
        CLIC running at 380\,GeV (left and right plots) and
        1.5\,TeV (right plot).}
          \label{fig:excluded}
\end{figure}

The expected limits on the invisible decays of the 125\,GeV Higgs boson
and limits on the production of new ``invisible'' scalars, which were
obtained in a model-independent approach, can also be used to
constrain various BSM scenarios.
We demonstrate the possibility of constraining parameters of the Higgs-portal models
taking the VFDM model~\cite{vfdm1,vfdm2} as an example.
The SM is extended by the spontaneously broken extra
$U(1)_X$ gauge symmetry and a Dirac fermion.
To generate mass for the dark vector $X_\mu$, the Higgs mechanism with
a complex singlet $S$ is used in the dark sector.
A new scalar state $\phi$, which describes a real-part fluctuation of $S$, 
can mix with the SM Higgs field $h$ implying the existence of
two mass eigenstates: 
\begin{equation*} 
 \left( 
\begin{array}{c}
H\\ 
H'
\end{array} 
\right) =
\left( 
\begin{array}{cc}
\cos \alpha   & \sin \alpha \\ 
-\sin \alpha &\cos \alpha
\end{array} 
\right)
\left(
\begin{array}{c}
h\\ 
\phi
\end{array} 
\right) \; ,
\end{equation*} 
where we assume that $H$ is the observed 125\,GeV state. 
If $\alpha \ll 1$, it is SM-like, but it can also decay invisibly (to
dark sector particles) via the $\phi$ component (BR$(H \to inv)\sim \sin^2\alpha$).  
If $H'$ is also light, it can be produced in e$^+$e$^-$ collisions
in the same way as the SM-like Higgs boson.
Limits on the mixing angle, $\sin\alpha$, resulting from
the cross section limits presented in Figure \ref{fig:limit}, are shown
in Figure \ref{fig:excluded}. 

\section{Conclusion}
We studied the sensitivity of  CLIC running at 380\,GeV and 1.5\,TeV
to invisible decays of the 125\,GeV Higgs boson and the possible production of new 
scalar states.
Associated production of a Higgs-like neutral scalar with a $Z$ boson  was considered.
The analysis based on the \whizard event generation and
fast simulation of the CLIC detector response with \delphes.
For 1000\,fb$^{-1}$ of data collected at 380\,GeV operation of CLIC, 
invisible Higgs boson decays
at the level of 1.0\% can be excluded at 95\% C.L.
Expected limits on the production cross section of the new scalar $H'$ were
presented as a function of its mass, for CLIC running at 380\,GeV  and  1.5\,TeV.

\section*{Acknowledgements}

The work was carried out in the framework of the CLIC detector and
physics (CLICdp) collaboration.
We thank collaboration members for fruitful discussions, valuable
comments and suggestions.
%
% \paragraph{Funding information}
The work was partially supported by the National Science Centre
(Poland) under OPUS research projects nos. 2017/25/B/ST2/00496
(2018-2021) and 2017/25/B/ST2/00191, and a HARMONIA project under 
contract UMO-2015/18/M/ST2/00518 (2016-2019).

\bibliography{dis2021_hinv.bib}

\begin{thebibliography}{10}
\providecommand{\url}[1]{\texttt{#1}}
\providecommand{\urlprefix}{URL }
\expandafter\ifx\csname urlstyle\endcsname\relax
  \providecommand{\doi}[1]{doi:\discretionary{}{}{}#1}\else
  \providecommand{\doi}{doi:\discretionary{}{}{}\begingroup
  \urlstyle{rm}\Url}\fi
\providecommand{\eprint}[2][]{\url{#2}}

\bibitem{ATLAS:2020cjb}
{The ATLAS Collaboration},
\newblock \emph{{Search for invisible Higgs boson decays with vector boson
  fusion signatures with the ATLAS detector using an integrated luminosity of
  139 fb$^{-1}$}},
\newblock ATLAS-CONF-2020-008 (2020).

\bibitem{CMS}
A.~M. Sirunyan \emph{et~al.},
\newblock \emph{{Search for invisible decays of a Higgs boson produced through
  vector boson fusion in proton-proton collisions at $\sqrt{s} =$ 13 TeV}},
\newblock Phys. Lett. \textbf{B793}, 520 (2019),
\newblock \doi{10.1016/j.physletb.2019.04.025},
\newblock \eprint{1809.05937}.

\bibitem{Mekala:2021ftq}
K.~Mekala, A.~F. Zarnecki, B.~Grzadkowski and M.~Iglicki,
\newblock \emph{{Sensitivity to invisible scalar decays at CLIC}},
\newblock Eur. Phys. J. Plus \textbf{136}(2), 160 (2021),
\newblock \doi{10.1140/epjp/s13360-021-01116-5}.

\bibitem{Thomson:2015jda}
M.~Thomson,
\newblock \emph{{Model-independent measurement of the e$^{{+}}$ e$^{-}$
  $\rightarrow $ HZ cross section at a future e$^{{+}}$ e$^{-}$ linear collider
  using hadronic Z decays}},
\newblock Eur. Phys. J. C \textbf{76}(2), 72 (2016),
\newblock \doi{10.1140/epjc/s10052-016-3911-5},
\newblock \eprint{1509.02853}.

\bibitem{Bambade:2019fyw}
P.~Bambade \emph{et~al.},
\newblock \emph{{The International Linear Collider: A Global Project}}  (2019),
\newblock \eprint{1903.01629}.

\bibitem{vfdm1}
A.~Ahmed, M.~Duch, B.~Grzadkowski and M.~Iglicki,
\newblock \emph{{Multi-Component Dark Matter: the vector and fermion case}},
\newblock Eur. Phys. J. \textbf{C78}(11), 905 (2018),
\newblock \doi{10.1140/epjc/s10052-018-6371-2},
\newblock \eprint{1710.01853}.

\bibitem{vfdm2}
M.~Iglicki,
\newblock \emph{{Vector-fermion dark matter}} (2018), \eprint{1804.10289}.

\bibitem{clicdet}
D.~Arominski \emph{et~al.},
\newblock \emph{{A detector for CLIC: main parameters and performance}},
\newblock CLICdp-Note-2018-005 (2018), \eprint{1812.07337}.

\bibitem{delph}
J.~de~Favereau \emph{et~al.},
\newblock \emph{{DELPHES 3, A modular framework for fast simulation of a
  generic collider experiment}},
\newblock JHEP \textbf{02}, 057 (2014),
\newblock \doi{10.1007/JHEP02(2014)057},
\newblock \eprint{1307.6346}.

\bibitem{delcards}
E.~Leogrande, P.~Roloff, U.~Schnoor and M.~Weber,
\newblock \emph{{A DELPHES card for the CLIC detector}}  (2019),
\newblock \eprint{1909.12728}.

\bibitem{whiz1}
W.~Kilian, T.~Ohl and J.~Reuter,
\newblock \emph{{WHIZARD: Simulating Multi-Particle Processes at LHC and ILC}},
\newblock Eur. Phys. J. \textbf{C71}, 1742 (2011),
\newblock \doi{10.1140/epjc/s10052-011-1742-y},
\newblock \eprint{0708.4233}.

\bibitem{whiz2}
M.~Moretti, T.~Ohl and J.~Reuter,
\newblock \emph{O'mega: An optimizing matrix element generator},
\newblock AIP Conference Proceedings \textbf{583}(1), 173 (2001),
\newblock \doi{10.1063/1.1405295}.

\bibitem{gHZZ}
A.~Robson and P.~Roloff,
\newblock \emph{{Updated CLIC luminosity staging baseline and Higgs coupling
  prospects}},
\newblock CLICdp-Note-2018-002 (2018), \eprint{1812.01644}.

\bibitem{Robson:2020lhl}
A.~Robson, P.~Roloff and J.~de~Blas,
\newblock \emph{{CLIC Higgs coupling prospects with a longer first energy
  stage}}  (2020),
\newblock \eprint{2001.05278}.

\bibitem{TMVA}
A.~Hocker \emph{et~al.},
\newblock \emph{{TMVA - Toolkit for Multivariate Data Analysis}},
\newblock CERN-OPEN-2007-007 (2007), \eprint{physics/0703039}.

\bibitem{Abbiendi:2002qp}
G.~Abbiendi \emph{et~al.},
\newblock \emph{{Decay mode independent searches for new scalar bosons with the
  OPAL detector at LEP}},
\newblock Eur. Phys. J. C \textbf{27}, 311 (2003),
\newblock \doi{10.1140/epjc/s2002-01115-1},
\newblock \eprint{hep-ex/0206022}.

\bibitem{Wang:2020lkq}
Y.~Wang, M.~Berggren and J.~List,
\newblock \emph{{ILD Benchmark: Search for Extra Scalars Produced in
  Association with a $Z$ boson at $\sqrt{s}=500$ GeV}}  (2020),
\newblock \eprint{2005.06265}.

\end{thebibliography}

\nolinenumbers

\end{document}